# A CONSUMER'S GUIDE TO LATTICE QCD RESULTS

T. DeGrand

*Department of Physics*
*University of Colorado*
*Boulder, Colorado 80309*

ABSTRACT: I present an overview of recent lattice QCD results on hadron spectroscopy and matrix elements. Case studies include light quark spectroscopy, the determination of $\alpha_s$ from heavy quark spectroscopy, the D-meson decay constant, a calculation of the Isgur-Wise function, and some examples of the (lack of) effect of sea quarks on matrix elements. The review is intended for the non-expert.

Talk presented at the 1993 Slac Summer Institute

# I. INTRODUCTION

This lecture is an introduction to lattice calculations in quantum chromodynamics for the non-expert "consumer." Lattice methods are presently the only way to perform calculations of masses and some matrix elements in the strong interactions beginning with the Lagrangian of QCD and including no additional parameters. By "consumer" I mean a person who might want to use a lattice calculation (as an input to a phenomenological calculation or to compare to her experiment, for example), and is not really interested in doing the lattice calculation herself, but would like to be able to judge the reliability of calculations in the literature.

There are many good reviews and introductions to lattice gauge theory and its use in QCD.[1] The lattice community has a large annual meeting and the proceedings of those meetings (Lattice 'XX, published so far by North Holland) are the best places to find the most recent results. However, as in any large community with its own set of problems, most of the papers in those proceedings tend to talk to each other in a language which is rather opaque to nonmembers. My goal is an impressionistic overview of the field as it presently exists, which might be useful to an outsider.

The bottom line is that for the past one or two years there have been a lot of lattice calculations of masses and matrix elements which agree with experiment at the ten to fifteen per cent level.

I will begin with a very superficial overview of how lattice calculations are performed. Then I will turn to a set of case studies: spectroscopy of light hadrons, of heavy quark systems, and of glueballs, then two case studies of matrix elements: the decay constants of D- and B-mesons, a recent calculation of the Isgur-Wise function, and some pictures of the effects of sea quarks on simple matrix elements.

## II. HOW LATTICE CALCULATIONS ARE CARRIED OUT

Lattice calculations are performed using the Euclidean path integral formulation of quantum field theory. If we have some field theory with field variables $\phi$ ($\phi$ could be quarks, gluons,...) and a Lagrange density $\mathcal{L}(\phi)$, we define an analog of the partition function in statistical mechanics

$$Z = \int [d\phi(x,t)] \exp(-\int d^4x \mathcal{L}(\phi)) \tag{1}$$

(here $x_\mu = (x, it)$). The expectation value of any observable $O(\phi)$ is given by

$$\langle O \rangle = \frac{1}{Z} \int [d\phi(x,t)] O(\phi) \exp(-\int d^4x \mathcal{L}(\phi)). \tag{2}$$

To be able to perform calculations in any quantum field theory one must introduce a short distance cutoff which regulates the ultraviolet divergences. We do that by replacing continuous space time by a lattice of grid points $x = ax_i$ where $a$ is the lattice spacing, and defining the field on those grid points $\phi(x) \to \phi_i = \phi(x_i)$. Then the functional integrals Eqns. (1) and (2) become ordinary integrals of very high dimensionality. One evaluates Eqn. (2) using importance sampling: somehow one creates an album of snapshots of the field variables $\phi_i$ where the probability that a particular configuration is present in the album is $P(\phi_j) = \exp(-\sum_x \mathcal{L}(\phi_j)$ and then

$$\langle O \rangle = \frac{1}{N} \sum_{j=1}^{N} O(\phi_j) + O(\frac{1}{\sqrt{N}}). \tag{3}$$

The generation of the album is done using Monte Carlo techniques not too different in principle from the ones you would use in an experiment to generate Monte Carlo events.

Lattice calculations are hard for several reasons:

1. The lattice spacing should be small–small enough that physics on a size scale less than a lattice spacing can be described using perturbation theory.

2. The size of the simulation volume $L^4$ should be greater than the physical size of the hadrons. This point is in conflict with item 1. The number of grid points is $n = (L/a)^4$. A gluon field is a three by three complex matrix per each direction on each lattice site, or 72 real numbers per lattice site. Fermions have four spins and three colors or 24 real numbers per site. Typical simulations have lattice spacings around 1/10 fermi (within a factor of two) and a number of mesh points ranging from $16^3 \times 32$ to $24^3 \times 40$ to $32^4$: the end is not yet in sight!

3. One needs a lot of statistics–tens to hundreds of uncorrelated lattice measurements.

4. It is very hard to compute with light (u,d) quark masses at their physical values. On the lattice calculating a quark propagator $G_q(x, x')$ involves inverting the matrix problem $(\slashed{D} - m)G_q(x, x') = \delta^4(x - x')$. The matrix becomes singular as $m_q \to 0$. One typically performs a calculation at an unphysical value of the light quark mass and then tries to extrapolate to $m_q = 0$.

5. Sea quarks are a problem because of Fermi statistics, which effectively introduces long range interactions among the quarks. There are techniques for dealing with this problem[2,3] but they make QCD with dynamical fermions orders of magnitude more difficult than if the sea quarks were not there (and the difficulty scales inversely as a power of the quark mass). A rather drastic approximation called the quenched approximation neglects this problem simply by throwing away all the sea quarks. This is an uncontrolled approximation which people do mainly because the alternative (keeping light sea quarks) is too time consuming for the computer.

All these constraints add up to a very hard numerical problem. We use the fastest supercomputers available. Cray's are usually too slow. Some groups have built their own computers. One of the projects I belong to used half of a Connection Machine CM-2 (at a speed of about 3 1/2 Gflops) for about two years. This is not considered an excessive amount of resources.

Finally, there are two more problems to watch out for.

6. In a lattice calculation all observables are measured on the same set of lattices and are highly correlated. There are methods for dealing with correlated data. Some lattice practitioners use them. My advice is that if the paper you are reading does not make some attempt to deal with the correlations which are present in its data (or is not aware that its data is correlated), you should discard the paper.

7. The major problem facing lattice calculations these days are systematics: Quenching, is $a$ small enough, is $L$ big enough, is the quark mass small enough? Lattice calculations produce as output not a hadron mass $m_H$ but the combination $am_H$. One finds $a$ by dividing $am_H$ by a measured $m_H$ (in MeV). The problem is, which mass to use? Most lattice calculations only reproduce mass ratios at the ten or fifteen per cent level, so the lattice spacing is uncertain at that level. This uncertainty propagates into essentially all interesting calculations.

I would be remiss if I did not provide you with a small glossary of lattice terms in order to enable you to read the literature:

1. Lattice people define $\beta = 6/g^2$ where $\alpha_s = g^2/4\pi$. Here $g$ is the color coupling constant measured at a momentum scale $Q \simeq \pi/a$, so bigger $\beta$ corresponds to smaller $a$.

2. "Link"– The vector potential $A_\mu(x)$ has an orientation and so instead of being defined on the sites of the lattice, is defined on the links joining adjacent points $x$ and $x + an_\mu$. For technical reasons lattice people use the "link variable" $U_\mu(x) = \exp(igaA_\mu(x))$ in simulations rather than the vector potential.

3. "Plaquette"– The lattice analog of the gauge action $F^2_{\mu\nu}(x)$ is the product of four links about a unit square or "plaquette" on the lattice.

4. "Staggered fermions"– On the lattice the quark energy momentum dispersion relation changes from its continuum value $E^2 = p^2 + m^2$ to $\sinh^2 Ea = \sin^2 pa + m^2 a^2$. This has low energy states near $pa = 0$ and degenerate extra solutions at the ends of the Brillouin zone $pa = \pi$. With staggered fermions these extra states are treated as extra spin or flavor degrees of freedom.

5. "Wilson fermions" add extra terms to the action to raise the energy of the $p = \pi/a$ modes and eliminate the extra degeneracy. The physical quark mass must be derived from the simulation in terms of an input parameter called the hopping parameter $\kappa$ and a measured parameter $\kappa_c$: $am_q = 1/2(1/\kappa - 1/\kappa_c)$. Wilson fermion calculators like to use $\kappa$ in their graphs.

## III. SPECTROSCOPY

All lattice calculations begin with spectroscopy. In order to measure the mass of a hadron which has some set of quantum numbers, invent an operator $J$ which has the same set of quantum numbers and compute

$$C(t) = \langle 0|J(t)J(0)|0\rangle. \tag{4}$$

Using the Euclidean version of the Heisenberg equation of motion

$$J(t) = \exp(Ht)J\exp(-Ht) \tag{5}$$

and inserting a complete set of energy eigenstates, we find

$$C(t) = \sum_n |\langle 0|J|n\rangle|^2 \exp(-E_n t) \tag{6}$$

which at big $t$ goes over to

$$C(t) \simeq |\langle 0|J|1\rangle|^2 \exp(-E_1 t) \tag{7}$$

where $E_1$ is the lightest state with the quantum numbers of $J$. The exponential falloff of the correlator gives us the mass, while its intercept gives us a matrix element $\langle 0|J|1\rangle$.

For bound states of quarks the operator $C(t)$ is basically the Feynman graph shown in Fig. 1: it is made of the appropriate number of quark and antiquark quarks propagating in the background of gluon fields in your album of configurations.

In the old days (pre-1988) the operators $J$ were local currents, like $\bar{\psi}\gamma_5\psi$ for the pion. Nowadays we use some big extended operator like $\sum_x \sum_y \phi(x,y)\bar{\psi}(x)\gamma_5\psi(y)$ which "looks like" a hadronic wave function. Then the computer has to do less work to filter out the lightest state. This means that Eqn. (6) takes its asymptotic form Eqn. (7) at a small $t$ value, while the signal is still large.

Spectroscopic studies in QCD involve light quarks, heavy quarks, and glueballs, so I will say a few words about each.

### Light Quarks

Generally in lattice calculations people try to deal with dimensionless quantities as much as possible, since they are independent of the precise value of the lattice spacing. In spectroscopy, people present their data on so-called "Edinburgh plots," $M_N/M_\rho$ vs. $M_\pi/M_\rho$ or "APE plot," $M_N/M_\rho$ vs. $(M_\pi/M_\rho)^2$. (The names are after the collaborations which invented the plots).

The most interesting recent quenched calculation is by a group from IBM which built its own computer to do QCD [4]. Fig. 2 shows their data plotted by me on an Edinburgh plot. There appears to be a small change between the data at larger lattice spacing ($\beta = 5.7$,

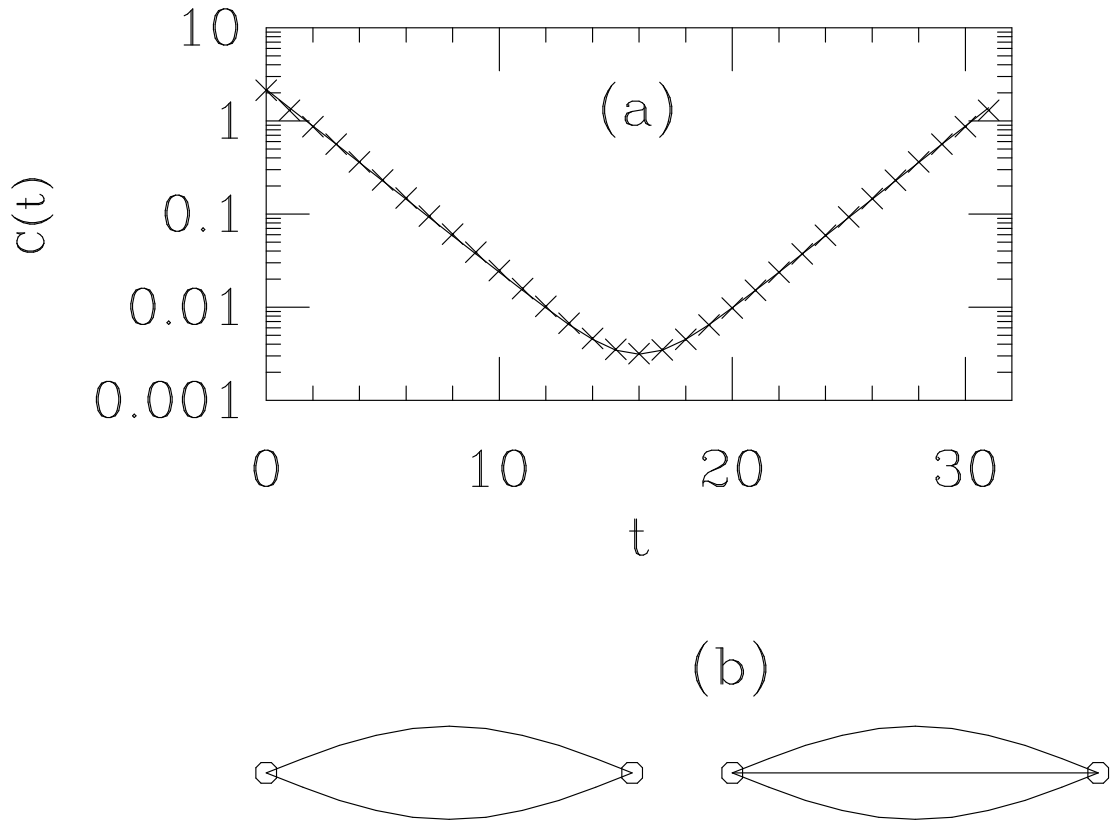

FIGURE 1

(a) A typical correlator showing good exponential falloff (the correlator has periodic boundary conditions in the time direction). (b) Feynman diagrams for meson and baryon correlators.

about 0.14 fm) to the smaller lattice spacing ($\beta = 5.93$, 6.17, $a$ down to about 0.07 fm). The authors of Ref. [4] have extrapolated their masses in $a$ and $L$ and present the limits in Fig. 3, as a plot of mass divided by $M_\rho$ at $M_\pi = 0$. The agreement with observation is spectacular.

All of these calculations are done in quenched approximation. Simulations with dynamical fermions are much more expensive and the data is correspondingly more meagre.

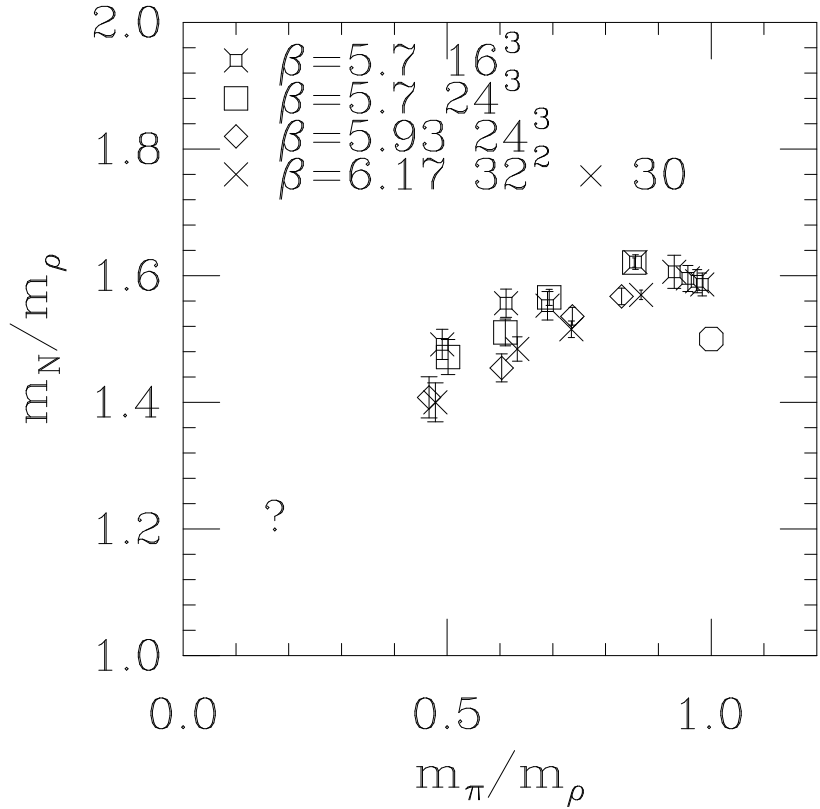

FIGURE 2

Edinburgh plot prepared by me from the data of Ref. 4, showing ratios at several values of the lattice spacing (different $\beta$'s). The octagon shows the expected result at infinite quark mass, and the question mark is the real world value.

As an example, I show In Fig. 4 an Edinburgh plot for simulations with two flavors of dynamical staggered fermions. I have connected the points with the same lattice spacing (same $\beta$). Again, the $N/\rho$ ratio appears to "settle" a bit as $\beta$ increases. The upper curve is an analytic calculation at infinitely strong coupling, $\beta = \infty$, where the lattice spacing is about 1/2 Fermi [5]. The lower curve is the extrapolation of the $\beta = 5.7$ data to zero quark mass. We see that in that limit the $N/\rho$ ratio is still too large. The lattice spacings here are all much larger than in Fig. 2–at $\beta = 5.445$ $a = 0.22$ fm, at $\beta = 5.6$ $a = 0.11$ fm, and at $\beta = 5.7$ $a = 0.089$ fm. (The relation between $\beta$ and the lattice spacing is different at $n_f = 0$ then at $n_f \neq 0$ because the QCD $\beta$-function depends on $n_f$.) These calculations have a ways to go.

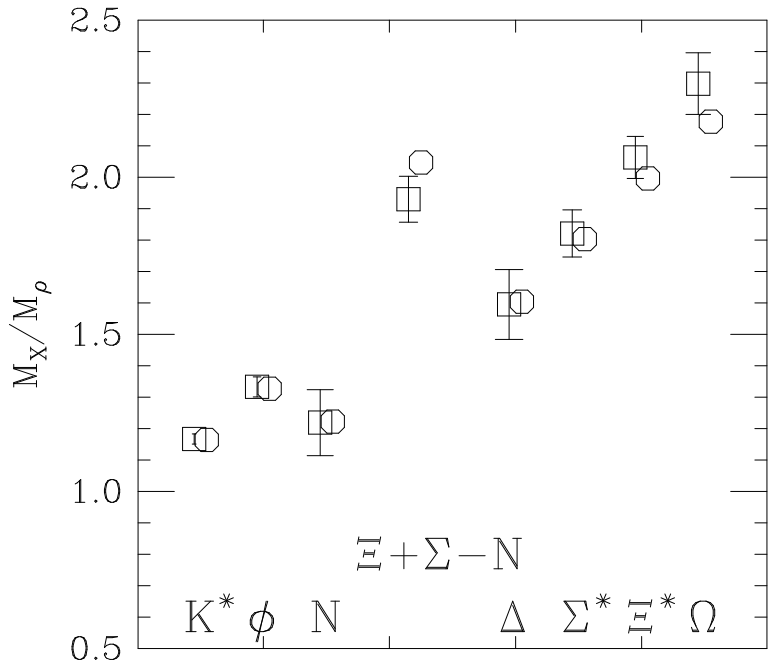

FIGURE 3

Ratio of lattice masses to the rho mass, after extrapolations to infinite simulation volume and zero lattice spacing, from Ref. 4. Circles are real world data, squares from simulations

## Heavy Quarks

People have been using the lattice to do calculations of heavy quark systems, too. As an example, Fig. 5 is a picture of the fine structure splitting in charmonium, by me and M. Hecht[10]. It looks just like the wallet card, although with large errors.

The most interesting application of heavy quark physics recently reported is the calculation of $\Lambda_{\overline{MS}}$ or $\alpha_{\overline{MS}}$. The idea is that the lattice needs one parameter to set the scale. The easiest parameter to determine on the lattice is a mass or mass difference and the idea is to use a mass difference to find $\Lambda_{\overline{MS}}$. The calculation has two parts. First, one must do a long distance calculation to find a mass; the lattice spacing comes from $a = M_H a / M_H(expt)$. Next, one must do a short distance calculation on the lattice to con-

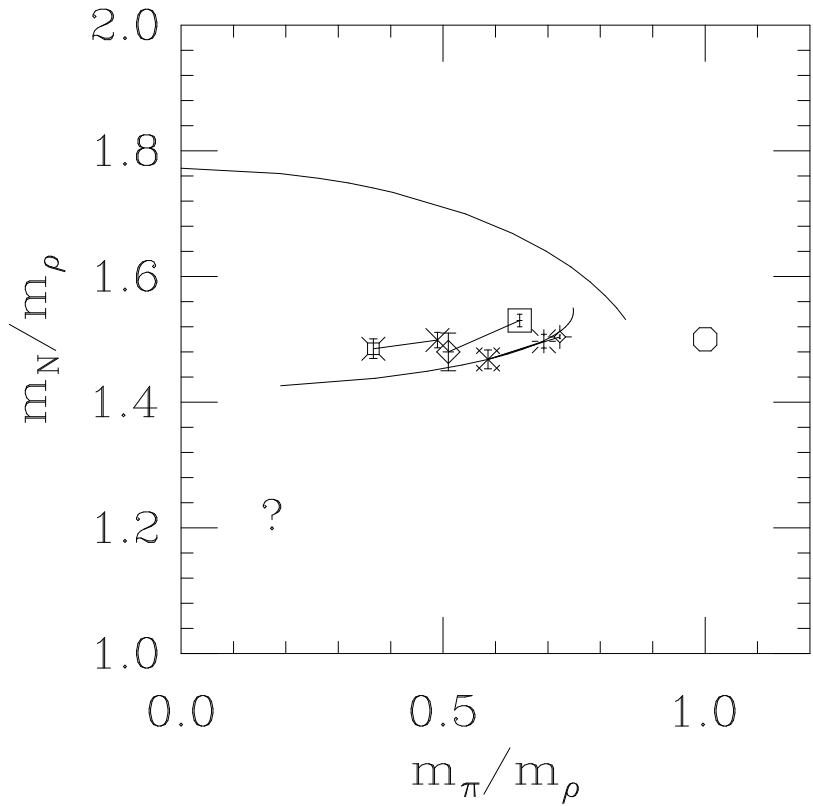

FIGURE 4

Edinburgh plot from simulations with two flavors of dynamical staggered fermions. Data are by Bernard. et. al.[6] (cross and fancy square), Bitar, et. al. [7] (square and diamond), F. Butler, et. al. [8] (fancy diamond) and M. Fukugita, et. al. [9](fancy cross and burst), while the upper curve is a theoretical prediction in strong coupling (from Ref. 5) and the lower curve is the extrapolation of the data at $\beta = 5.7$ to zero quark mass.

nect something perturbative to lattice perturbation theory and then to a continuum ($\overline{MS}$) number. The calculation has been done by two groups: El-Khadra, Hockney, Kronfeld, and Mackenzie[11], and Davies, Lepage and Thacker[12].

The calculation begins by noticing that the mass difference between the lightest S-wave and P-wave $Q\bar{Q}$ mesons is nearly independent of quark mass (in the $\psi$ system it is 460 MeV, in the $\Upsilon$, 430 MeV. Since this difference is independent of the quark mass, one does not have to tune the quark mass on the lattice in order to measure it. So the lattice S-P mass splitting gives the lattice spacing.

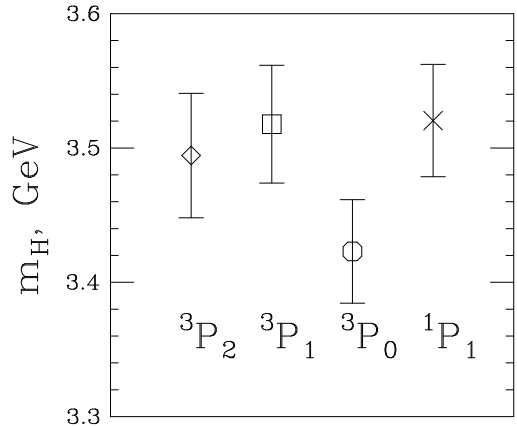

FIGURE 5

Masses of the four P-wave $c\bar{c}$ states, from Ref. 10.

Next, one must determine the coupling constant. This is done through the short range $Q\bar{Q}$ potential,

$$V(q^2) = \frac{4}{3}\alpha_s(q^2)\frac{1}{q^2} \tag{8}$$

where $q \simeq \pi/a$. Lattice perturbation theory does a good job at the lattice spacings of the simulation, and so one can measure the short distance potential, extract $\alpha_s$ on the lattice, and carry out the conversion. Doing so gives $\alpha_{\overline{MS}}^{n_f=0}(5\text{GeV}) = 0.140(4)$.

The main problem with the calculation as it presently stands is that it is done in quenched approximation. One must somehow convert $\alpha_{\overline{MS}}^{n_f=0}$ to $\alpha_{\overline{MS}}^{n_f=4}$. The authors of Ref. [11] do this by running the coupling constant down from the upsilon mass to the typical bound state $Q$ with no flavors, then out with four flavors. They find that this shifts $\alpha_s$ by $25 \pm 6$ per cent–that is, the uncertainty in the amount of the shift is itself 25 per cent. This seems conservative. The bottom line is that $\alpha_{\overline{MS}}^{n_f=4}(5\text{ GeV}) = 0.174(12)$. A second, completely separate calculation using nonrelativistic quarks [12] gives $\alpha_{\overline{MS}}^{n_f=4}(5\text{ GeV}) = 0.170(10)$. Finally, running down to the Z-mass gives

$$\alpha_{\overline{MS}}^{n_f=4}(M_Z) = 0.105(4) \tag{9}$$

This result is shown with other determinations of $\alpha_s$ in Fig. 7. It is a little bit lower than most of the results from LEP.

At the conference, several theorists wanted to make a big stir that the lattice number is somehow better than the other determinations, and that it should be trusted more than

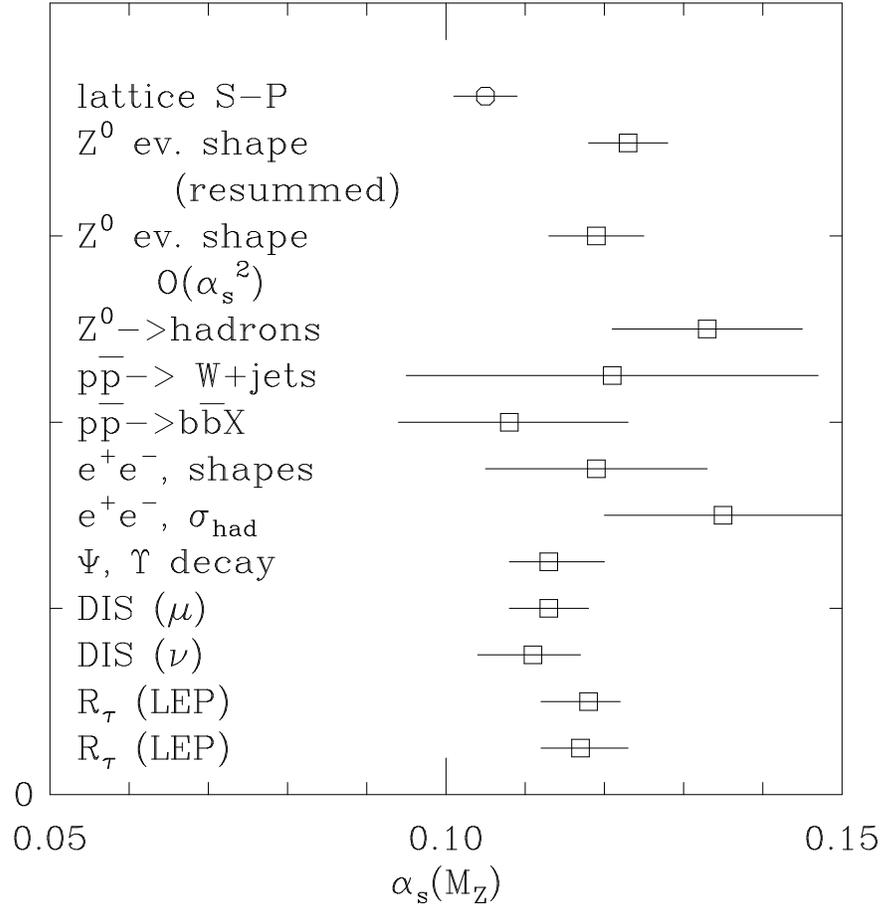

FIGURE 6

The strong coupling constant at the mass of the Z. The lattice number is shown along with other determinations from various experiments, from Ref. 13.

the other ones. While I feel that the lattice number will eventually turn out to be more reliable, since it only uses spectroscopy as an input, no jet physics, I don't think that is the case yet. I would really prefer to see checks at smaller lattice spacing and (eventually) the full lattice simulation done in the presence of dynamical fermions. Maybe I just don't know where all the bodies are buried in the $e^+e^-$ analyses!

Heavy quark systems are better than light quark systems for exploring the effects of quenching, because light quarks only modify the potential between the heavy quarks. One can play with models to understand their effects. In contrast, we don't really understand the effects of sea quarks on light hadron spectroscopy. For example, why are the rho and omega mesons nearly degenerate even though the decay width of the rho is so much greater than the omega's?

FIGURE 7

Glueball masses from the UKQCD collaboration [14] The black symbols are measurements and the open symbols are just lower limits on the masses. (This figure is Fig. 1 of Ref. 14.)

Glueballs

People have been calculating glueball masses in QCD for many years. Unfortunately, all the high statistics calculations use quenched approximation; you would find nothing interesting in any of the calculations which include sea quarks. A representative recent compilation[14] is shown in Fig. 7. All of these states are above $\pi\pi$ threshold and so one would worry how these numbers will change when sea quarks are included. That's an open problem.

## IV. MATRIX ELEMENTS

Most of the matrix elements measured on the lattice are expectation values of local operators composed of quark and gluon fields. The mechanical part of the lattice calculation begins by writing down some Green's function which contains the local operator (call it $J(x)$) and somehow extracting the matrix element. For example, if one wanted $\langle 0|J(x)|h\rangle$ one could look at the two-point function

$$C_{JO}(t) = \sum_x \langle 0|J(x,t)O(0,0)|0\rangle \qquad (10).$$

Inserting a complete set of correctly normalized momentum eigenstates

$$1 = \frac{1}{L^3}\sum_{A,\vec{p}} \frac{|A,\vec{p}\rangle\langle A,\vec{p}|}{2E_A(p)} \qquad (11)$$

and using translational invariance and going to large $t$ gives

$$C_{JO}(t) = e^{-m_A t}\frac{\langle 0|J|A\rangle\langle A|O|0\rangle}{2m_A}. \qquad (12)$$

A second calculation of

$$C_{OO}(t) = \sum_x \langle 0|O(x,t)O(0,0)|0\rangle = e^{-m_A t}\frac{|\langle 0|O|A\rangle|^2}{2m_A} \qquad (13)$$

can be used to extract $\langle 0|J|A\rangle$ (fit two correlators with three parameters).

Similarly, a matrix element $\langle h|J|h'\rangle$ can be gotten from

$$C_{AB}(t,t') = \sum_x \langle 0|O_A(t)J(x,t')O_B(0)|0\rangle. \qquad (14)$$

(Can you see how?)

These lattice matrix elements are not yet the continuum matrix elements. The lattice is a UV regulator and changing from the lattice cutoff to a continuum regulator (like $\overline{MS}$) introduces a shift

$$\langle f|O^{cont}(\mu=1/a)|i\rangle_{\overline{MS}} = a^D(1+\frac{\alpha_s}{4\pi}(C_{\overline{MS}} - C_{latt}) + \ldots)\langle f|O^{latt}(a)|i\rangle + O(a) + \ldots. \qquad (15)$$

The factor $a^D$ converts the dimensionless lattice number to its continuum result. The $O(a)$ corrections arise because the lattice operator might not be the continuum operator: $df/dx = (f(x+a) - f(x))/a + O(a)$. The C's are calculable in perturbation theory. There are a number of tricks/deep theoretical ideas for achieving a more convergent perturbation expansion.[15]

## A Case Study–Heavy Meson Decay Constants

The decay constant $f_M$ of a pseudoscalar meson $M$ is defined as

$$\langle 0|\bar{\psi}\gamma_0\gamma_5\psi|M\rangle = f_M m_M. \tag{16}$$

Decay constants are interesting because some of them ($\pi$ and $K$) are measured and provide a benchmark for lattice calculations, while some of them are not measured and allow predictions ($D$, $D_s$, and $B$). They probe very simple properties of the wave function: in the nonrelativistic quark model

$$f_M = \frac{\psi(0)}{\sqrt{m_M}} \tag{17}$$

where $\psi(0)$ is the $\bar{q}q$ wave function at the origin. For a heavy quark ($Q$) light quark ($q$) system $\psi(0)$ should become independent of the heavy quark's mass as the $Q$ mass goes to infinity, and in that limit one can show in QCD that $\sqrt{m_M}f_M$ approaches a constant. It is believed that CP nonconserving amplitudes are proportional to $f_M^2$ and so knowledge of $f_B$ provides information about CP nonconservation in the B system.

One way to compute the decay constant is to put a light quark and a heavy quark on the lattice and let them propagate. It is difficult to calculate $f_B$ directly on present day lattices because the lattice spacing is much greater than the b quark's Compton wavelength (or the UV cutoff is below $m_b$). In this limit the b quark is strongly affected by lattice artifacts as it propagates. However, one can make $m_b$ infinite on the lattice and determine the combination $\sqrt{m_B}f_B$ in the limit. Then one can extrapolate down to the $B$ mass and see if the two extrapolations up and down give the same result. Until a year or so ago the two methods did not give consistent numbers. However, the present situation is that one can reliably compute $f_D$ and $f_B$ in quenched approximation.

As an example, results from a recent calculation by Bernard, Labrenz, and Soni[16] is shown in Fig. 8. What is plotted is $f_M\sqrt{M}/(1+\alpha_s/\pi \log(Ma))$ vs. $1/M$; the extra term is a perturbative correction to the static heavy quark formula. Removing it allows one to interpolate to infinite quark mass. Their predictions are reproduced in Table I. There are several other lattice predictions of these numbers. (See Ref. 17 for a compilation.) They differ in detail, but all give numbers in the range of those of Table I. There are two experimental measurements of $f_{D_s}$. They are 232±45±20±48 MeV [18] or 344±37±52±42 MeV[19]. The error bars are too big for a serious comparison.

## The Isgur-Wise Function

The physics of systems containing a heavy quark and a light quark has a very simple limit as the mass of the heavy quark goes to infinity. The physics of the light quark becomes independent of the mass or other properties of the heavy quark. (For an extensive review, see Ref. 20.) In particular, the form factor in $B \to D$ semileptonic decay is described by a universal function called the Isgur-Wise function $\xi_0(v'\cdot v,\mu)$ which depends on the four-velocities of the two heavy quarks:

$$\langle D(v')|\bar{c}\gamma_\nu b|B(v)\rangle = \sqrt{m_B m_D}C_{cb}(\mu)\xi_0(v'\cdot v,\mu)(v+v')_\nu \tag{18}$$

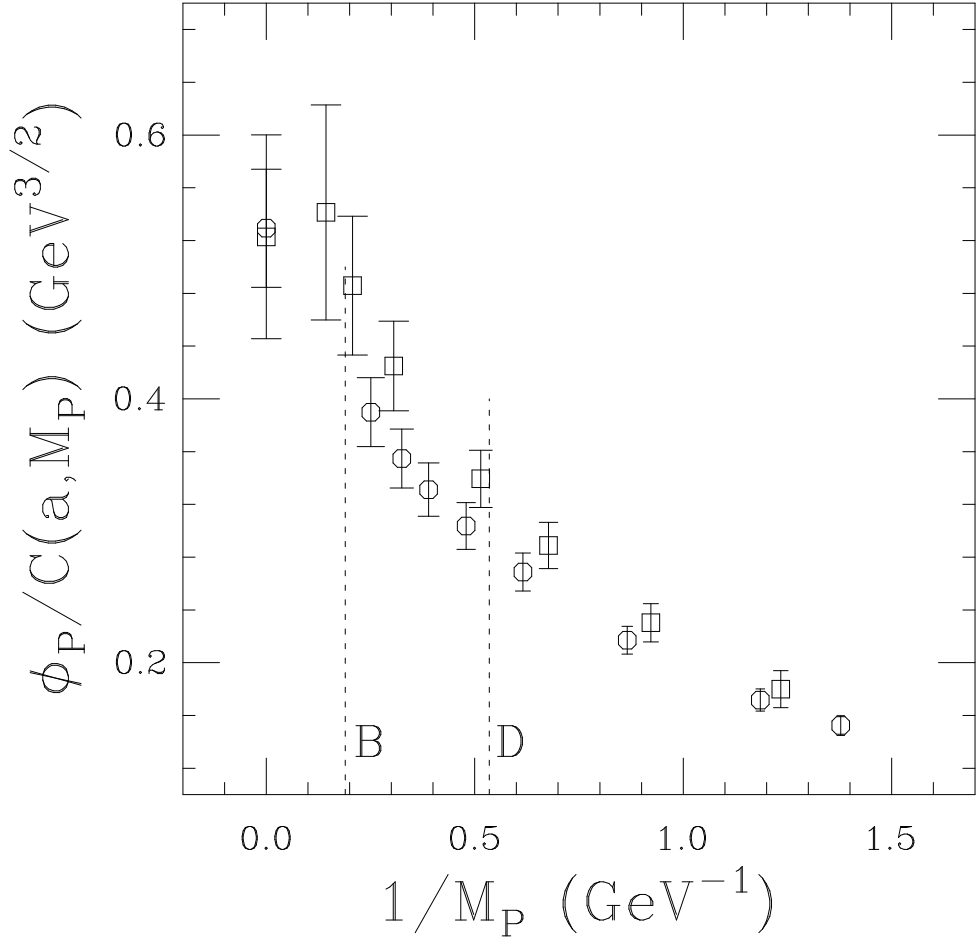

FIGURE 8

$f_M \sqrt{M}/(1 + \alpha_s/\pi \log(ma))$ vs. $1/M$, from Ref. 16. The results from two lattice spacings ($\beta$ values) are shown, to give a rough idea of lattice spacing systematics.

($C_{bc}$ is a short-distance perturbative factor). One would like to calculate $\xi_0(v \cdot v')$ from first principles. While there has been some discussion of how to do this with infinite mass heavy quarks on the lattice[21], another technique is just to calculate a form factor on the lattice and fit it to the form of Eqn. (18). Bernard, Shen, and Soni[22] have recently published a preprint which does just that, by measuring the form factor

$$\langle D(v')|\bar{c}\gamma_\nu c|D(v)\rangle = m_D C_{cc}(\mu)\xi_0(v' \cdot v, \mu)(v + v')_\nu. \tag{19}$$

The lattice calculation spans the range $0 < v \cdot v' < 1.2$ while real-world data ranges over $1.1 < v \cdot v' < 1.5$.

Their results are shown in Figs. 9 and 10. Fig. 9 shows lattice $\xi_0(v \cdot v')$ as a function of $(v \cdot v')$ for a variety of light and heavy quark masses (see Ref. 22 for details) and also shows various theoretical predictions: an upper bound on the Isgur-Wise function [23] (dashed line), a fit to the lattice data (solid line), and one theoretical prediction of the curve [24] (dotted curve). Fig. 10 shows a comparison with the experimental data from ARGUS[25].

Table I. Predictions for heavy-light decay constants from Ref. 16, showing various uncertainties.

| particle | $f_P$, $MeV$ | fitting/extrap | scale | large-$am$ |
|---|---|---|---|---|
| $B$ | 187(10) | ±12 | ±15 | ±32 |
| $B_s$ | 207(9) | ±10 | ±22 | ±32 |
| $D$ | 208(9) | ±11 | ±12 | ±33 |
| $D_s$ | 230(8) | ±10 | ±18 | ±28 |

Of course this is just the beginning of these calculations. The lattice can be used to test the universality of the Isgur-Wise function at the B and D-meson masses, and determine the dependence of corrections to it on the heavy quark mass.

## Testing the Quenched Approximation

All of these calculation are performed in the quenched approximation. There is an unknown systematic associated with throwing away the sea quarks. The only way I know to really test it is to repeat the simulations with dynamical sea quarks. That is very expensive. However, there are a few tests already in the market. Fig. 11 shows the decay constant of vector mesons parameterized by

$$\langle V | V_\mu | 0 \rangle = \frac{1}{f_V} m_V^2 \epsilon_\mu. \qquad (20)$$

from a simulation with Wilson valence quarks and two flavors of dynamical staggered quarks [26]. The two plotting symbols are for two different values of the sea quark mass (the lattice spacing is about $1/a \simeq 2$ GeV). Clearly the effects of sea quarks are small. As a second example, we display the pseudoscalar decay constant from the same data set against the results of Ref. 16 in Fig. 12. Our data (the burst) is at about the same lattice spacing as the quenched data plotted as squares. If there is an effect of sea quarks, it is not very large.

FIGURE 9

Lattice data from various combinations of heavy and light quark masses from Ref. 22 with some theoretical curves superposed. (This figure is Fig. 1 of Ref. 22.)

## SUMMARY

Present day lattice calculations are able to produce ten to fifteen per cent numbers for a wide variety of physical observables. Most of the uncertainties are systematics limited (at the cost of large amounts of computing to beat down statistics). The major systematic is the lattice spacing. It is just not understood how small the lattice spacing should be so that lattice calculations are insensitive to it (or more precisely, so that all physics on scales less than $a$ are perturbative). There are claims[27] that heavy quark spectroscopy only needs $a \simeq 1/5$ fm. However, glueball spectroscopy at that lattice spacing shows $a$ dependence. The D-meson decay constant needs $a \simeq 0.08$ fm. The minimum lattice spacing is probably process dependent.

FIGURE 10

Argus data for the $B \to C$ form factor from Ref. 25 in circles, with the lattice results of Ref. 22 as crosses. (This figure is Fig. 3 of Ref. 22.)

Simulations remain unwieldy. Doing almost anything requires at least a year of work on a supercomputer. This project length seems to be an invariant–as computers improve, our standards have gone up. Nevertheless, the continued improvement in computer hardware allows us to tackle more and more complicated (interesting?) projects, so that the field will continue to advance even in the absence of new ideas.

The most interesting new ideas, which might lead to improved calculations on smaller computers, are concerned with the question: Can one find a more complicated discretization which allows one to work at bigger lattice spacings? Doubling the number of terms in the lattice action roughly doubles the amount of work, while halving the lattice spacing at fixed simulation volume increases the work by a factor of 16. This subject is under active study.[28]

The best way to end the talk is to quote a previous speaker (Prof. B. Frois): "You

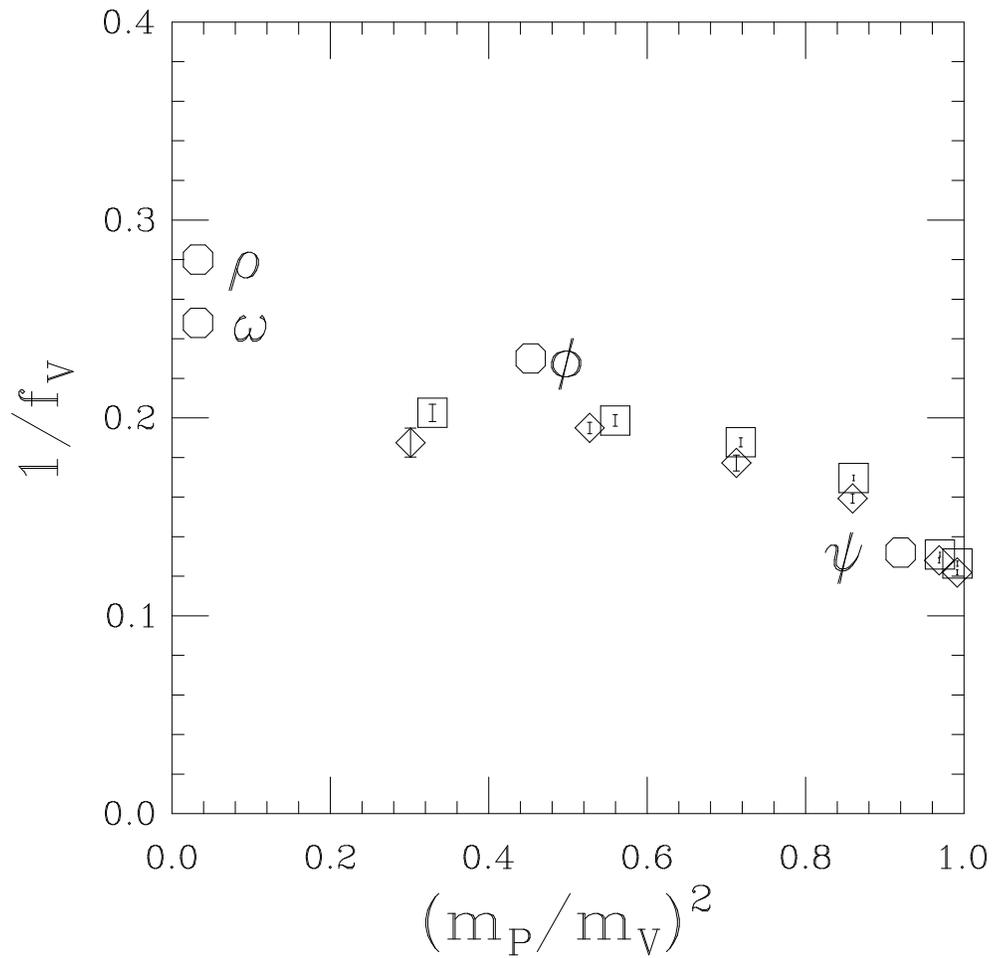

FIGURE 11

Lattice $1/f_V$ from Ref. 26. The labeled points are physical particles Results from simulations with sea quark mass in lattice units $am_q = 0.01$ are shown in squares, and for sea quark mass $am_q = 0.025$ in diamonds.

have to fight your way to the numbers."

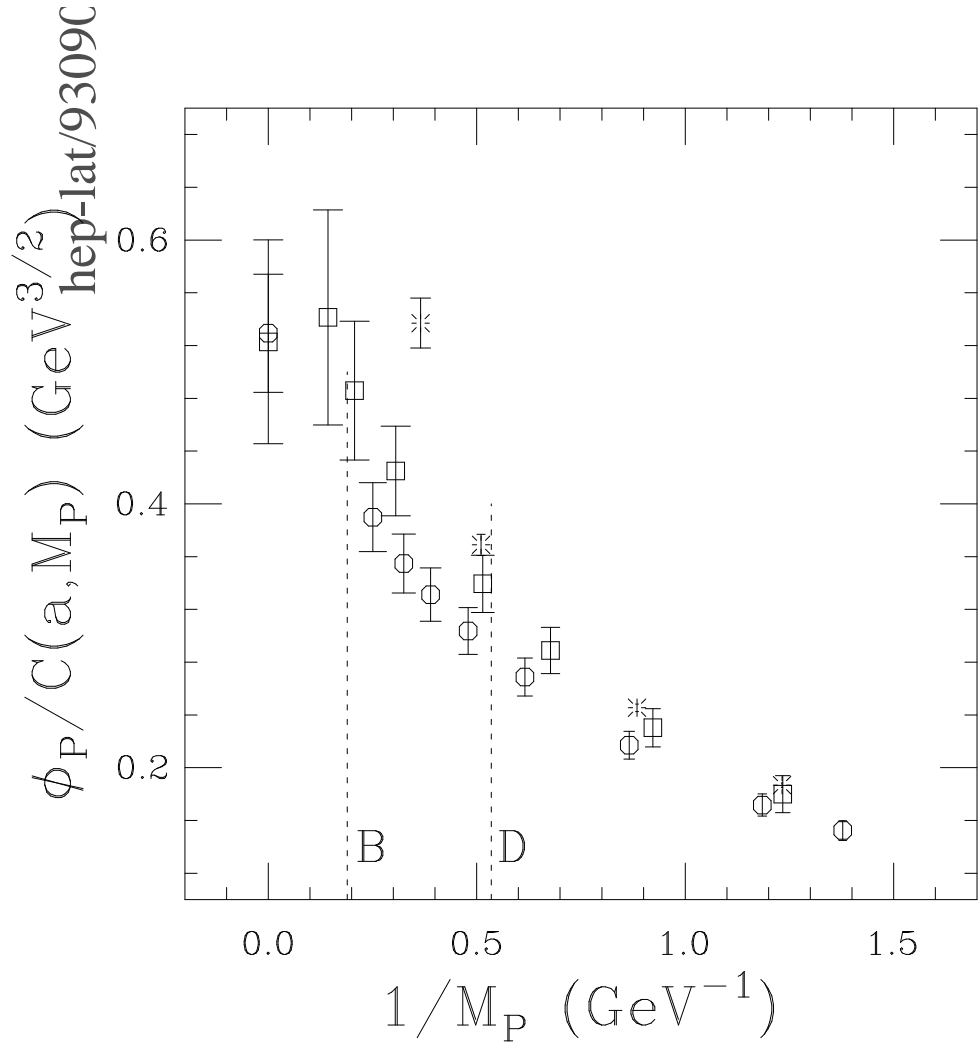

FIGURE 12

Pseudoscalar decay constants from Ref. 16 (quenched) and from Ref. 26 (burst), at sea quark mass $am_q = 0.01$ and equivalent lattice spacing.

## ACKNOWLEDGMENTS

I would like to thank Jim Labrenz, A. Soni, and Don Weingarten for providing me with their data and for discussions about it, and the conference organizers for putting on a very useful meeting. I would also like to thank the clapper rail for standing still. This work was supported by the U.S. Department of Energy.